\documentstyle[eqsecnum,aps]{revtex}
\newcommand{\vc}[1]{{\bf {#1}}}
\begin{document}
\draft
\title{Raman cooling of atoms below the gravitational limit}
\author{A. V. Soroko\footnote{Electronic address: soroko@hep.by}}
\address{National Center of Particle and High Energy Physics,
Belarusian State University, \\
Bogdanovich Street 153, Minsk 220040, Belarus}
%\date{\today}
\maketitle
\begin{abstract}
Raman  cooling   of  non-zero-spin   atoms  in   the  presence  of
gravitational and external magnetic  fields is investigated.   The
magnetic  field  is   adjusted  so  as   to  compensate  for   the
gravitational force acting on ground-state atoms.  The dark  state
(DS) is created  and supported in  momentum space with  additional
velocity-selective two-photon  transitions.   The minimum  allowed
temperature  is  found  to  be  determined  only  by  the width of
velocity  selection  and  therefore  can  be  much  less  than the
gravitational  limit.    A  complete  set  of  analytical formulas
describing cooling  of a  dilute atomic  sample is  derived.  They
serve as the basis for numerical simulations which are carried out
in the one-dimensional (1D) case.
\end{abstract}

%\narrowtext

% body of paper here
\section{INTRODUCTION}
\label{Introduction}

Methods of laser cooling of atoms have progressed dramatically  in
recent  years.    There  are  three  typical  temperature   scales
characterizing various methods.  The first one is specific to  the
most common  scheme in  which Doppler  shift causes  the radiation
pressure  force  to  be  velocity  dependent,  thus damping atomic
motion  when  the  laser  frequency  is  tuned  below  an   atomic
resonance.  The minimum temperature $T_D$ for atoms cooled in such
a way is known as the "Doppler limit".  It is proportional to  the
natural   width   of   the   laser-driven   transition    $\gamma$
\cite{Phillips}, $k_B T_D= \hbar\gamma/2$, and for the $D$ line of
Na the Doppler limit is approximately 240 $\mu$K.

Schemes based on dissipation of atomic energy via interaction with
the vacuum modes  of electromagnetic field  have a lower  limit on
the achievable temperature defined by the minimum of recoil energy
which  an  atom  obtains  after  spontaneous photon emission.  The
corresponding scale is known as the recoil limit $k_B T_R = (\hbar
k)^2/2M$, where $k$ is the  wave number of emitted light,  and $M$
is the atomic mass.  For Na it approximately equals to 1 $\mu$K.

To overcome  this limit  two subrecoil  cooling methods  have been
developed   and   demonstrated:      velocity  selective  coherent
population  trapping  (VSCPT)   \cite{VSCPT}  and  Raman   cooling
\cite{Raman}.  Both methods  imply the existence of  the so-called
dark  state,  which  does  not  interact  with  light,  has a long
lifetime  and  occupies  only  a  few  modes  in  momentum   space
\cite{Dum}.    During  the  cooling  cycle atoms diffuse into this
state due  to random  recoil induced  by spontaneous  emission and
accumulate in  it.   Since DS  has a  vanishing absorption rate of
light, the final temperature is restricted not by the recoil limit
but by the time of cooling which, however, cannot be greater  than
the lifetime of DS.

In the  absence of  Earth gravity,  infinitely long  cooling times
would be  possible.   In practice  the gravitational  field pushes
atoms from  DS, reducing  its lifetime  dramatically.   For any 3D
configuration this  defines the  third characteristic  temperature
scale \cite{Dum2} $k_B T_G =  Mg/k$, the gravitational limit.   It
lies below the recoil limit for most atoms, e.g., $T_G =  0.07T_R$
for Rb, and $T_G = 0.003T_R$ for Na.

Two  ways  are  envisioned  to  prepare  a stable quantum state of
matter  in  the  gravitational  field:    to bound particles or to
suspend  them  free  in  an  inhomogeneous  magnetic  field  using
Stern-Gerlach effect.  In the first approach atoms are confined by
a conservative trapping potential which can be realized, e.g.,  in
a far-off-resonance or a dipole trap, where an intensity  gradient
provides a spatially dependent ac Stark shift.  In momentum space,
up to now only the existence of an approximate dark state has been
demonstrated \cite{Pellizzari}, characterized by a decay rate in a
special 1D atomic and laser field configuration much smaller  than
that of  all other  states in  the trap.   The  finite lifetime of
approximate DS evidently restricts the cooling possibilities in  a
trap, leaving  the question  about going  below the  gravitational
limit to be clarified.   However, a scheme \cite{Morigi} which  is
based on the creation of a  dark state in position space with  the
help of an appropriate spatial profile of the cooling laser, e.g.,
in a doughnut mode, seems  to be much more efficient,  allowing to
cool a significant  fraction of atoms  to the ground  state of the
trapping potential.

Another approach  may be  applied to  atoms possessing  a magnetic
moment.  Superimposing a weakly inhomogeneous magnetic field  onto
the path  of pre-polarized  particles and  appropriately adjusting
the field  gradient it  is possible  to compensate  the effects of
gravity  for  a  definite  internal  atomic  state.   However, the
magnetic field  induces spatially  dependent shifts  of the Zeeman
levels, which lead to unwanted residual excitation from the DS  in
the  framework  of  any  traditional  subrecoil  cooling   method.
Moreover,  in  the  case  of  VSCPT  the  dark  state cannot be an
eigenvector  of  the  total  Hamiltonian  since  only  one  of the
internal states forming the superposition which is not coupled  to
the laser field  may escape gravity.   Thus, both  VSCPT and Raman
cooling mechanisms  in their  standard form  are incompatible with
the last approach.

To resolve this problem we suggest a modification of Raman cooling
method, in which the  ground-level atoms are made  motionally free
with the Stern-Gerlach effect and the DS is created and  supported
in    momentum    space    of    these   atoms   with   additional
velocity-selective two-photon transitions.  The transitions couple
external momentum states of the same ground internal level and are
organized in such a  manner that DS cyclically  occupies different
thin sets of  velocity modes while  remaining unreachable for  the
Raman excitation-repumping pulse sequences at all times.

In  Sec.\  \ref{sec2}  a  detailed  qualitative  treatment  of the
suggested scheme is given.  For reasonable experimental conditions
all the stages of  the scheme admit analytical  descriptions which
are   presented   in   Sec.\   \ref{sec3}   -   Sec.\  \ref{sec5}.
Specifically,  the  formulas  describing  a  coherent   two-photon
transition when an atom is placed in a superposition of two  plain
electromagnetic  waves  with  arbitrary  directions  of  the  wave
vectors are presented  in Sec.\ \ref{sec3}.   In Sec.\  \ref{sec4}
Raman excitation to the closest hyperfine level is investigated in
the  regime  in  which  the  photon  spontaneous  emission  may be
neglected.  Both the  exact quadrature and convenient  approximate
expressions are derived.  In Sec.\ \ref{sec5} the optical  pumping
of atoms back to the ground state is considered for short times of
light  pulse,  i.e.,  when  the  external potential field does not
affect the atomic motion substantially.  A numerical simulation of
100  cooling  sequences  in  one  dimension  is  given  in   Sec.\
\ref{sec6}.  Section  \ref{sec7} concludes with  a summary of  the
obtained results.

\section{QUALITATIVE TREATMENT}
\label{sec2}

Consider  for  definiteness  an  atom  with  a  $J=\case{1}{2}$ to
$J=\case{3}{2}$ transition, e.g., sodium or cesium.  The  magnetic
field  $\vc{B}(\vc{r})$  applied  to  compensate  the  gravity  is
supposed to  contain a  homogeneous component  $\vc{B}_0$ directed
along the gravity acceleration $\vc{B}_0 \uparrow\uparrow \vc{g}$.
The remaining inhomogeneous part of the field  $\vc{B}_{1}(\vc{r})
=  \vc{B}(\vc{r})  -  \vc{B}_0$  should  be small compared to this
component,
\begin{equation}
|\vc{B}_{1}(\vc{r})| \ll B_0 = |\vc{B}_0|.
\label{1.0}
\end{equation}
As we will see below, to fulfil this condition it is necessary  to
take $B_0$ in  the range $10^3  \div 10^4$ G.  In practice such  a
field is strong enough to induce Zeeman shifts which  considerably
exceed the  hyperfine splitting  intervals $\sim  \hbar\omega_{\rm
HFS}$ (but not the multiplet ones).  Therefore an internal  atomic
eigenstate $|J,I,M_J,m_I \rangle$ may be well described using  the
set of quantum  numbers consisting of  the angular momenta  of the
electronic  shell  $J$  and  the  nucleus  $I$,  and  their  local
projections $M_J$, $m_I$ on the direction of the magnetic field.

In the  framework of  perturbation theory,  $|J,I,M_J,m_I \rangle$
represents    a    combination    of   eigenstates   $|J,I,M_J,m_I
\rangle^{(0)}$  related  to  the  atomic  Hamiltonian  without the
hyperfine interaction,
\begin{eqnarray}
|J,I,M_J,m_I  \rangle =&&
|J,I,M_J,m_I  \rangle^{(0)} + \frac{a}{2\mu_B g_L B_0}
\nonumber \\
&&\times \left\{ [(J+M_J)(J-M_J+1)]^{1/2}
[(I+m_I+1)(I-m_I)]^{1/2} \right.
\nonumber \\
&&\times |J,I,M_J-1,m_I+1  \rangle^{(0)}
\nonumber \\
&&-[(J+M_J+1)(J-M_J)]^{1/2}
[(I+m_I)(I-m_I+1)]^{1/2}
\nonumber \\
&&\times \left.|J,I,M_J+1,m_I-1  \rangle^{(0)} \right\},
\label{eq1}
\end{eqnarray}
where   $a$   is   the   hyperfine  coupling  constant  ($a\propto
\hbar\omega_{\rm HFS}$, e.g.,  for Na $a/\hbar  = 885.8$ MHz)  and
$g_{L}$  denotes  the  Lande  factor.    The  corresponding energy
eigenvalue is determined not only by the multiplet level $E_J$ but
also  by  the  magnetic  field  $B(\vc{r}) = |\vc{B}(\vc{r})|$ and
therefore is spatially dependent
\begin{eqnarray}
E_{|J,I,M_J,m_I\rangle}(\vc{r}) = &&
E_J + a M_J m_I \nonumber \\
&& + (\mu_B g_L M_J - \mu_{\rm nuc} m_I) B(\vc{r}),
\label{eq2}
\end{eqnarray}
where $\mu_{\rm nuc}$ is the nuclear magnetic moment.  Because  of
the  condition  (\ref{1.0})  such  a  spatial dependence, however,
mainly arises from the longitudinal ($B^{\parallel}_{1}(\vc{r})  =
\vc{B}_0  \cdot   \vc{B}_{1}(\vc{r})  /B_0$),   rather  than   the
transverse ($\vc{B}^{\perp}_{1}(\vc{r})$) component of the  vector
$\vc{B}_{1}(\vc{r})$,  provided  that  the  components are defined
relative to $\vc{B}_0$.  This is evident from the expression
\begin{eqnarray}
B(\vc{r}) = &&
\sqrt{\left[B_0 + B^{\parallel}_{1}(\vc{r})\right]^2
+ \left[\vc{B}^{\perp}_{1}(\vc{r})\right]^2} \nonumber \\
&& \simeq B_0 + B^{\parallel}_{1}(\vc{r}) +
\left[\vc{B}^{\perp}_{1}(\vc{r})\right]^2/(2 B_0),
\label{1.3}
\end{eqnarray}
where the  term containing  $\vc{B}^{\perp}_{1}(\vc{r})$ is  small
and can be neglected.  Consequently, by adjusting the gradient  of
the    field    $B^{\parallel}_{1}(\vc{r})$    one   can   achieve
translational  invariance  of   the  ground  state   $|1\rangle  =
|1/2,I,-1/2,I \rangle$ in three dimensions:
\begin{equation}
E_{|1\rangle}(\vc{r}) - M\vc{g}\cdot\vc{r}=\mbox{\rm const}.
\label{eq3}
\end{equation}
For example, to  balance the gravitational  force in this  way for
sodium   it   is   necessary   to   create   a   gradient  $\nabla
B^{\parallel}_{1}(\vc{r})  =  b_1  \vc{g}/|\vc{g}|$,  where $b_1 =
-4.033$ G$/$cm.   This condition does  not contradict the  Maxwell
equation $\nabla \cdot \vc{B}_{1}(\vc{r}) = 0$, because  variation
of $\vc{B}^{\perp}_{1}(\vc{r})$ is not restricted.  Note also that
the  choice  $B_0  =  10^3  \div  10^4$  G maintains the condition
(\ref{1.0}) very well  within a spatial  region of the  size $\sim
10$ cm.

All  the  other  levels  are  affected  by  the  residual external
potential.  In particular, after a transition from $|1\rangle$  to
the neighboring  state $|2\rangle  = |1/2,I,-1/2,I-1  \rangle$ the
atom experiences a force
\begin{equation}
\vc{f}_2 =  \frac {M \mu_{\rm nuc}\vc{g}}
{(1/2)\mu_B g_L + \mu_{\rm nuc}I}.
\label{eq4}
\end{equation}

In  our  scheme,  we  use  pulses  of  laser  light at frequencies
$\omega_1$  and  $\omega_2$  which   are  roughly  tuned  to   the
$|1\rangle   \to   |3\rangle$   and   $|2\rangle   \to  |3\rangle$
transitions,  where  $|3\rangle  =  |3/2,I,-3/2,I  \rangle$  is an
excited state with  the lowest energy.   The typical  size $2L$ of
atomic sample is restricted by the condition $L \ll a/(Mg)$, which
allows  to  regard  $E_{|3\rangle}  (\vc{r})$  as  the  closest to
resonance  excited  level  within  the  whole  interaction domain.
Indeed, the  force $\vc{f}_3$  acting on  the atoms  in the  state
$|3\rangle$   may   be   estimated   from  Eqs.\  (\ref{eq2})  and
(\ref{eq3}) as $|\vc{f}_3| \sim Mg$. The maximal spatial shift  of
the  level  $\sim  MgL$  which  it  induces  is much less than the
hyperfine splitting intervals ($MgL  \ll a \sim \hbar  \omega_{\rm
HFS}$), and the hierarchy of detunings is retained.  Therefore  an
atom initially in  $|1\rangle$ or $|2\rangle$  state behaves as  a
three-level system with respect  to the processes with  stimulated
emission of photons.

Since the atomic dipole momentum operator $\hat\vc{d}$ is diagonal
in  quantum  numbers  $I$  and  $m_I$  in  the basis $|J,I,M_J,m_I
\rangle^{(0)}$,  the   transitions  which   change  $m_I$,   e.g.,
$|2\rangle  \to  |3\rangle$,  are  allowed  only  due to hyperfine
interaction, as is seen from  Eq.\ (\ref{eq1}).  The value  of any
matrix  element  like  $|\langle  3|  \hat\vc{d}  |2  \rangle|$ is
approximately  $\propto  \eta_{\rm  HF}  |\langle  3|   \hat\vc{d}
|1\rangle|$, where $\eta_{\rm HF} = a/(2\mu_B g_L B_0) \ll 1$.  As
a consequence,  the upper  state $|3\rangle$  decays to  the lower
ones preferentially in the channel $|3\rangle \to |1\rangle$ (with
the rate $\gamma$).  This  circumstance makes it possible to  deal
with an atom  as a three-level  system even if  spontaneous photon
emission takes place.

When the atom  is irradiated with  two laser beams  at frequencies
$\omega_1$ and  $\omega_2$, the  two-photon Raman  transition from
$|1\rangle \to |2\rangle$ has  twice the Doppler sensitivity  of a
single-photon transition provided  that $\omega_1 -  \omega_2 \sim
\omega_{\rm HFS}$ and the beam wave vectors $\vc{k}_1$, $\vc{k}_2$
are opposite \cite{Raman}.  However,  if we take into account  the
force  (\ref{eq4}),  a  wide  set  of  atomic momenta $\vc{p}$ may
satisfy  the  resonance  condition,  as  follows  from  the energy
conservation:
\begin{equation}
\hbar\Delta_1   -  2\vc{p}\cdot\vc{\Delta}_p/M
= \hbar\Delta_2   -  \vc{f}_2\cdot\vc{r}
+ 2\Delta_p^2/M.
\label{eq5}
\end{equation}
Here detunings $\Delta_{m} \equiv \omega_{m} + [E_{|m\rangle}(0) -
E_{|3\rangle}(0)]/\hbar$, $m=1,2$,  are defined  in the  center of
atom-laser  interaction  region  ($\vc{r}=0$),  $\vc{\Delta}_p   =
\hbar(\vc{k}_1  -  \vc{k}_2)/2$,  and  $\Delta_p=|\vc{\Delta}_p|$.
The dip in the velocity dependence of absorption rate broadens  so
that the width of the trapping zone \cite{Reichel} becomes
\begin{equation}
\delta v \sim L |\vc{f}_2|  /(2\Delta_p).
\label{tzwidth}
\end{equation}
As  a  consequence,  since  the  sample  of  unconfined  particles
considered in this paper may spread up to $L \sim 1$ cm during the
cooling,  the  effective  temperature  of  atoms left in the state
$|1\rangle$,  which  constitutes  $\sim  M  (\delta v)^2/(2 k_B)$,
generally lies far above the gravitational limit.  For example, in
the  case  of  sodium,  where  $\Delta_p/\hbar = 1.07 \times 10^5$
cm$^{-1}$    and    $|\vc{f}_2|/\hbar    =    7.3   \times   10^4$
cm$^{-1}$\,s$^{-1}$, such a temperature may reach $6.4 T_G$.

Despite insufficient  velocity selectivity  of the  $|1\rangle \to
|2\rangle$  transition,  state  $|2\rangle$  may  be used in Raman
excitation  cycle.    To  avoid  unwanted  radiation impact on the
selected group of particles, which  are referenced here as the  DS
atoms, one should  move them in  momentum space to  another place,
where the resonance condition (\ref{eq5}) brakes down.  It can  be
achieved  by  means  of  a  two-photon  $|1\rangle  \to |1\rangle$
transition while the atom is irradiated with two noncolinear laser
beams at the same frequency $\omega_1$.

\begin{figure}
\epsfbox{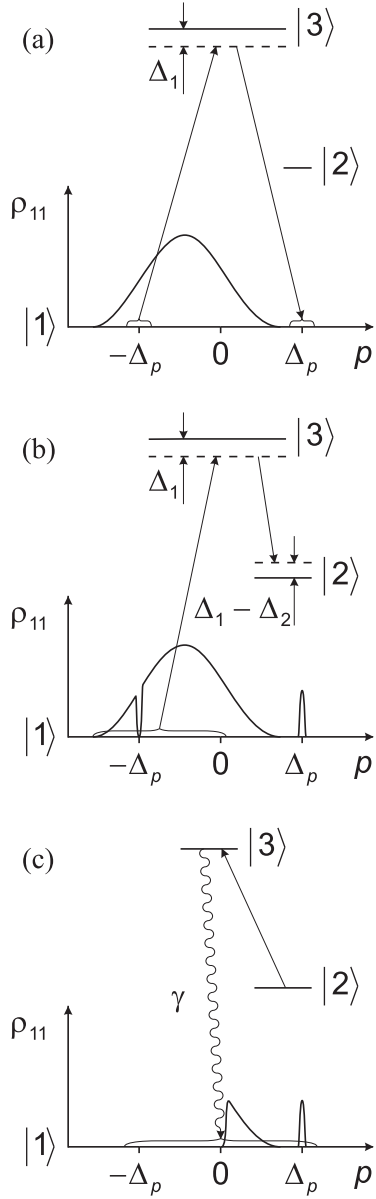}
\caption{
Energy  level   diagram  and   profile  of   ground  state  atomic
distribution  $\rho_{11}$  as  a  function  of  $p$,  the momentum
component  along  the  direction  of  vector $\vc{\Delta}_p$:  (a)
before DS transfer with  the two-photon $|1\rangle \to  |1\rangle$
transition;  (b)  before  the  Raman  excitation  $|1\rangle   \to
|2\rangle$ cycle; and (c) before the optical repumping pulse tuned
to the  $|2\rangle \to  |3\rangle$ transition.   The  curly braces
denote momentum intervals involved in each of these processes.
}
\label{fig1}
\end{figure}

If  the  ground-level  initial  momentum  distribution  along  the
direction  of  vector  $\vc{\Delta}_p$  were  as  shown  in  Fig.\
\ref{fig1}(a), such  a transition  would have  selectively brought
particles concentrated near the  point $-\Delta_p$ (the DS,  as we
will see below) to the point $\Delta_p$, and vice versa.  To prove
this  imagine  an  atom  with  momentum $\vc{p}$ passing through a
superposition  of  two  laser  beams.    The  superposition may be
treated  as   a  diffraction   grating  in   the  case   $\vc{k}_1
\uparrow\downarrow        \vc{k}_2$        (standing         wave)
\cite{Moskowitz,Martin}, or as  an effective atomic  hologram when
directions of the  wave vectors are  arbitrary \cite{Soroko}.   At
low laser light intensity  and large detuning $\Delta_1$  only the
first-order Bragg  scattering is  of importance  \cite{Kazantsev}.
In  this  case,  two  diffraction  modes  with  indices  0  and  1
resonantly   couple   with   each   other    \cite{Martin,Zhang2}.
Physically,  the  first-order  Bragg  resonance  corresponds to an
absorption and stimulated photon  emission process from one  laser
beam to another.   As a  consequence of the  atomic kinetic energy
conservation one gets the Bragg resonance condition
\begin{equation}
\pm\vc{p}\cdot\vc{\Delta}_p = \Delta_p^2,
\label{eq6}
\end{equation}
which  is   satisfied  for   any  momentum   with  the   component
$p=\pm\Delta_p$  along   the  vector   $\vc{\Delta}_p$.     Figure
\ref{fig1}(b)  contains  the  final  distribution, the peak around
$\Delta_p$ being the  moved DS.   So the {\it  first step} of  our
scheme consists in the momentum transfer of DS as it is  indicated
with arrows in Fig.\ \ref{fig1}(a).

In the {\it  second step} of  cooling, the Raman  excitation cycle
\cite{Raman} takes  place.   In accordance  with Eq.\ (\ref{eq5}),
atoms  with  any  negative   $p$  can  be  transferred   to  state
$|2\rangle$ by varying the difference of beam frequencies.  Due to
the finite width of trapping zone atoms with positive $p < M\delta
v$ also have a chance to undergo transition.  The DS, being hidden
near the  point $p=\Delta_p  > M\delta  v$, does  not take part in
this process, as illustrated in Fig.\ \ref{fig1}(b).

In the  {\it third  step}, an  optical pumping  pulse at frequency
$\omega_2$  is  used  to  return  the  atoms  back  to  the  state
$|1\rangle$.  It is important that the ground level appears to  be
far off  resonance and  laser light  does not  affect DS directly.
The  population  of  DS  rises  during  the  spontaneous  emission
process,   which   randomizes   the   atomic  momenta  [see  Fig.\
\ref{fig1}(c)].

Then the sequence of {\it steps 1 -- 3} is repeated with  opposite
directions  of  $\vc{k}_1$   and  $\vc{k}_2$  involving   residual
positive-momentum  atoms  of  the  ground  level in DS filling and
finishing a 1D  cooling cycle along  $\vc{\Delta}_p$.  After  this
stage  the  DS   occupies  its  initial   place  near  the   point
$-\Delta_p$.

By choosing  linearly independent  vectors $\vc{\Delta}_p^m$  in a
set of two  ($m=1,2$) or three  ($m=1,2,3$) 1D cooling  cycles one
can  proceed  with  decreasing  the  temperature  in  two or three
dimensions by repeatedly applying such sets.

To increase  the efficiency  of DS  filling one  can admit several
Raman and optical pumping pulses,  i.e., a number of {\it  steps 2
and 3},  between two  consecutive {\it  first steps}.   It  can be
done, for example, as in the classical method \cite{Raman},  where
every Raman transition is followed by the optical repumping, or by
applying a series of cycles, each including multiple Raman and one
optical pumping pulses.

Since  the  time  necessary  to  collect  all  the atoms in DS is,
generally speaking, infinitely long, it may be useful to  separate
the DS from background with the final {\it first-step} transitions
(on one for  each dimension) so  that the DS  and background atoms
will move in opposite directions and eventually will not spatially
overlap.      In   particular,   when  vectors  $\vc{\Delta}_p^m$,
$m=1,2,3$, form  an orthogonal  basis, our  scheme will  produce a
cooled atomic  beam with  the average  momentum $\vc{\Delta}_p^1 +
\vc{\Delta}_p^2   +   \vc{\Delta}_p^3$   as   follows   from  Eq.\
(\ref{eq6}).    The  minimum  allowed  temperature  (but  not  the
intensity) of such a beam is obviously determined by the width  of
velocity selection  specific to  {\it first-step}  transitions and
therefore can be much less than the gravitational limit.

\section{GROUND-STATE TWO-PHOTON TRANSITIONS}
\label{sec3}

In    contrast    to     the    case    of     Bragg    scattering
\cite{Martin,Kazantsev}, where diffracted modes are assumed to  be
spatially  resolvable  at  some  distance  from the light standing
wave, the present paper deals with short interaction times when an
atom moves  inside a  superposition of  two laser  beams from  the
beginning to the end.  This allows one to represent each beam as a
plane electromagnetic wave ($m=1,2$)
\begin{equation}
\vc{E}_{m}(\vc{r},t)=
\vc{E}_{m}
\exp(i\vc{k}_{m}\cdot\vc{r} - i\omega_{m} t) + c.c.,
\label{3.1}
\end{equation}
where  $\vc{E}_{m}$   stands  for   the  complex   amplitude,  and
$\omega_{2} = \omega_{1}$.

To simplify  the consideration  the coherent  scattering processes
are assumed to dominate the spontaneous emission, i.e., the regime
$|\Delta_{1}| \gg \gamma$ is kept \cite{Korsunsky}.  Under such  a
condition the one-particle density matrix \cite{Kazantsev} has  an
obvious time evolution
\begin{eqnarray}
\rho_{ab}(\vc{p}_1,\vc{p}_2,t)= &&
\int d \vc{p}'_1 \int d \vc{p}'_2
\sum_{a'b'} G_{aa'}(\vc{p}_1,\vc{p}'_1,t)
\nonumber\\
&& \times G^{*}_{bb'}(\vc{p}_2,\vc{p}'_2,t)
\rho_{a'b'}(\vc{p}'_1,\vc{p}'_2,t=0),
\label{3.2}
\end{eqnarray}
where indices $a,b  \dots$ denote the  internal atomic states  and
$G_{aa'}(\vc{p}_1,\vc{p}'_1,t)$  is  the  Green  function  of  the
two-component Shr\"{o}dinger  equation describing  atomic dynamics
during the $|1\rangle \leftrightarrow |3\rangle$ transitions.

In the rotating wave approximation the equation for slowly varying
in    time    ground-    and    excited-level    wave    functions
$\psi_{1}(\vc{p},t)$ and $\psi_{3}(\vc{p},t)$ takes the form
\begin{mathletters}
\begin{eqnarray}
i\frac{\partial}{\partial t} \psi_{1}(\vc{p},t) = &&
[t(\vc{p})+\Delta_{1}]\psi_{1}(\vc{p},t)
-\Omega^{*}_{1} \psi_{3}(\vc{p}+\hbar\vc{k}_{1},t) \nonumber\\
&&-\Omega^{*}_{2} \psi_{3}(\vc{p}+\hbar\vc{k}_{2},t),
\label{3.3a}
\end{eqnarray}
\begin{eqnarray}
i\frac{\partial}{\partial t} \psi_{3}(\vc{p},t) = &&
[t(\vc{p})-i\vc{f}_3\cdot\nabla]\psi_{3}(\vc{p},t)
-\Omega_{1} \psi_{1}(\vc{p}-\hbar\vc{k}_{1},t) \nonumber\\
&&-\Omega_{2} \psi_{1}(\vc{p}-\hbar\vc{k}_{2},t),
\label{3.3b}
\end{eqnarray}
\end{mathletters}
where $\Omega_{m} = \langle 3|\hat\vc{d}\cdot \vc{E}_{m}|1\rangle/
\hbar$,  $m=1,2$,  are  the   Rabi  frequencies,  and  the   terms
$t(\vc{p})=\vc{p}^2/(2M\hbar)$  and  $-i\vc{f}_3\cdot\nabla$ arise
in  momentum   space  from   the  kinetic   and  potential  energy
($-\vc{f}_3\cdot\vc{r}$) correspondingly.

For  the  situation  at  hand,  the  upper electronic state can be
adiabatically  eliminated  from  Eqs.\  (\ref{3.3a}), (\ref{3.3b})
provided   that   the   detuning   $\Delta_{1}$  is  large  enough
\cite{Martin,Zhang,Guzman}
\begin{equation}
|\Delta_{1}| \gg |\Omega_{1}|,|\Omega_{2}|,|\vc{f}_3|L/\hbar.
\label{condition1}
\end{equation}
The  route  by  which  one  can  do  it  implies a self-consistent
assumption $|\psi_{3}| \ll  |\psi_{1}|$ leading to  the zero-order
solution  of  the  Eq.\  (\ref{3.3a}):  $\psi_{1}(\vc{p},t) \simeq
\exp\{-i[t(\vc{p}) + \Delta_{1}] t\} \psi_{1}(\vc{p},t=0)$.  After
substitution of this expression into Eq.\ (\ref{3.3b}) the  latter
may be solved  in the framework  of perturbation theory  developed
with respect  to the  potential energy  term.   In this  case, the
excited-level wave function acquires a representation
\begin{eqnarray}
\psi_{3}(\vc{p},t)\simeq &&
\frac{\Omega_{1} \psi_{1}(\vc{p}-\hbar\vc{k}_{1},t)}
{t(\vc{p})-t(\vc{p}-\hbar\vc{k}_{1})-\Delta_{1}}
\nonumber \\
&&+\frac{\Omega_{2} \psi_{1}(\vc{p}-\hbar\vc{k}_{2},t)}
{t(\vc{p})-t(\vc{p}-\hbar\vc{k}_{2})-\Delta_{1}}
+ \ldots,
\label{3.4}
\end{eqnarray}
where  the  dots  denote  omitted  terms  which  include  a  small
($\propto |\vc{f}_3|L/|\hbar \Delta_{1}|$) first-order  correction
to $\psi_{3}(\vc{p},t)$ and also summands which oscillate with the
non-resonant frequency $t(\vc{p})$ and therefore give a negligible
contribution when one uses the above expression within the context
of Eq.\ (\ref{3.3a}).

For an ultracold atomic sample one can further neglect the kinetic
energy terms in the denominators of Eq.\ (\ref{3.4}) so that after
introducing of a new set of functions ($n\in Z$)
\begin{equation}
\psi^{(n)}(\vc{p},t) \equiv
\psi_{1}(\vc{p}+(2n-1)\vc{\Delta}_{p},t),
\label{3.5}
\end{equation}
Eq.\  (\ref{3.3a})  becomes  equivalent  to  an infinite system of
equations   defined    in   the    domain   ${\cal    D}=\{\vc{p}:
|\vc{p}\cdot\vc{\Delta}_p| \le \Delta_p^2\}$:
\begin{eqnarray}
i\frac{\partial}{\partial t} \psi^{(n)}(\vc{p},t) = &&
t^{(n)}(\vc{p})\psi^{(n)}(\vc{p},t)
\nonumber\\
&&+g\psi^{(n+1)}(\vc{p},t)
+g^{*}\psi^{(n-1)}(\vc{p},t),
\label{3.6}
\end{eqnarray}
where
\begin{equation}
t^{(n)}(\vc{p})  =
t(\vc{p}  +(2n-1)\vc{\Delta}_{p}) +\Delta_{1}
+\frac{|\Omega_{1}|^2}{\Delta_{1}}
+\frac{|\Omega_{2}|^2}{\Delta_{1}},
\label{3.7}
\end{equation}
and   $g=\Omega^{*}_{1}\Omega_{2}/\Delta_{1}$   stands   for   the
effective Rabi frequency.

At low $|g|$ only  the two functions with  $n=0$ and $n=1$ have  a
possibility  to  influence  each  other  resonantly  in the system
(\ref{3.6}) because  only $t^{(0)}(\vc{p})$  and $t^{(1)}(\vc{p})$
may be equal when $\vc{p} \in {\cal D}$.  If we take into  account
the coupling of  other functions, all  $\psi^{(n)}(\vc{p},t)$ will
get    corresponding    corrections    $\propto   \delta^{(n)}   =
|g|/(t^{(n)}(\vc{p})-  t^{(m)}(\vc{p}))$,  where  $m=n\pm  1$, and
$m+n\ne  1$.    Therefore  it  is  possible  to truncate relations
(\ref{3.6}), having  in mind  that $\delta^{(n)}  \ll 1$  when the
effective  Rabi  frequency  is  small  enough.   In this case, the
equations   for   $\psi^{(n)}(\vc{p},t)$   with   $n=0,1$   become
homomorphic with the rate  equations describing a two-level  atom,
and their  solution is  well known  (see, e.g.,  \cite{Letokhov}).
The  remaining  non-resonance  functions  simply  undergo  a  free
evolution.

However,  as   a  general   rule,  the   original  wave   function
$\psi_{1}(\vc{p},t)$ reconstructed in accordance with the  formula
(\ref{3.5})  appears   to  be   discontinuous  along   the  planes
$\vc{p}\cdot\vc{\Delta}_p =  0, \pm  2\Delta_p^2$.   To recover  a
smooth behavior, one  can modify the  reconstruction prescription,
e.g.,
\begin{eqnarray}
\psi_{1}(\vc{p},t)=&&
\psi^{(0)}(\vc{p}+\vc{\Delta}_{p},t) +
\psi^{(1)}(\vc{p}-\vc{\Delta}_{p},t)
\nonumber \\
&&- \exp[-i t^{(0)}(\vc{p}+\vc{\Delta}_{p}) t]
\psi_{1}(\vc{p},t=0),
\label{3.9}
\end{eqnarray}
where  the  solutions  $\psi^{(0,1)}(\vc{p},t)$  of  the truncated
system (\ref{3.6}) must be  analytically continued into the  whole
momentum space.   It  is easy  to check  that $\psi_{1}(\vc{p},t)$
built in such  a way coincides  with the exact  representation via
$\psi^{(n)}(\vc{p},t)$  up  to  an  error  of  order $\delta^{(0)}
\simeq   \delta^{(1)}   \simeq   |g|/t(2   \vc{\Delta}_{p})$  and,
consequently,   obeys   (with   the   same   accuracy)  the  Eqs.\
(\ref{3.3a}) -- (\ref{3.3b}).

As a result  the ground-state component  of the Green  function is
given by
\begin{eqnarray}
G_{11}(\vc{p},\vc{p}',t) =&&
\sum_{\sigma=0,\pm 1}
\phi^{(\sigma)}(\vc{p},t)
\delta^{3}(\vc{p}+2\sigma\vc{\Delta}_{p} - \vc{p}')
\nonumber\\
&&\times
\exp[i\omega_{1}t
-(i/\hbar)  E_{|3\rangle}(0)t],
\label{3.10}
\end{eqnarray}
where the following notations are used:
\begin{equation}
\phi^{(\pm 1)}(\vc{p},t)=
\frac{-i g^{(\pm 1)} } {d^{(\pm 1)}}
\exp(-i b^{(\pm 1)} t) \sin(d^{(\pm 1)} t),
\label{3.11}
\end{equation}
\begin{eqnarray}
\phi^{(0)}(\vc{p},t) =&&
\sum_{\sigma=\pm 1}
\left[
\frac{i a^{(\sigma)}}
{d^{(\sigma)}}
\sin(d^{(\sigma)} t)
+ \cos(d^{(\sigma)} t)
\right]
\nonumber\\
&&\times\exp(-i b^{(\sigma)} t)
\nonumber\\
&&- \exp[-i t^{(0)}(\vc{p}+ \vc{\Delta}_{p}) t].
\label{3.12}
\end{eqnarray}
In these formulas $g^{(+1)}= g$, $g^{(-1)}= g^{*}$,
\[
a^{(\pm 1)}= [t(\vc{p}\pm 2\vc{\Delta}_{p}) -t(\vc{p})]/2,
\]
\[
b^{(\pm 1)}=a^{(\pm 1)} + t^{(0)}(\vc{p}+ \vc{\Delta}_{p}),
\]
and
\[
d^{(\pm 1)}= \sqrt{(a^{(\pm 1)})^2+|g|^2}.
\]

It is seen  from Eqs.\ (\ref{3.10})  -- (\ref{3.12}) that  an atom
with  an  initial  momentum  component  (along  $\vc{\Delta}_{p}$)
$p\simeq\Delta_p$ will change it  to $p\simeq-\Delta_p$ at a  time
$\tau_n$ (the time of the $n\pi$ pulse)
\begin{equation}
\tau_n=\frac{\pi}{2|g|}(2n+1), \quad n\in Z.
\label{3.13}
\end{equation}
This transition is velocity-selective with the maximum  efficiency
determined  by  the  Bragg  resonance  condition (\ref{eq6}).  The
width of the peak in momentum distribution (the interval from  the
maximum to the first minimum) depends on the interaction time  and
for $t=\tau_0$ is
\begin{equation}
\delta p = \sqrt{3}\hbar |g| M/\Delta_p.
\label{3.14}
\end{equation}
For a  given $\Delta_p$  it decreases  with $|g|$.   Therefore one
should use a  large detuning and  small Rabi frequencies  to get a
narrower peak.

\section{RAMAN EXCITATION}
\label{sec4}

Consider  a  three-level  atom  placed  in  the field of two plane
electromagnetic  waves  (\ref{3.1})  with  different   frequencies
($\omega_1 -  \omega_2 \sim  \omega_{\rm HFS}$).   As  before, the
regime  of  large  detunings  is expected ($|\Delta_{\{1,2\}}| \gg
\gamma$), which allows to neglect spontaneous emission and to  use
Eq.\ (\ref{3.2}) for finding the density matrix evolution provided
that $G_{aa'}(\vc{p}_1, \vc{p}'_1,t)$ is interpreted as the  Green
function of the three-component Shr\"{o}dinger equation describing
atomic dynamics during the $|1\rangle \leftrightarrow  |3\rangle$,
$|2\rangle \leftrightarrow |3\rangle$ transitions.

If the  detunings are  also large  enough in  comparison with  the
maximal spatial  shifts of  transition frequencies  and the  laser
intensities are far below saturation, i.e., the conditions
\begin{equation}
|\Delta_{2}|  \gg  |\Omega_{1}|,|\Omega_{2}|,
|\vc{f}_3-\vc{f}_2|L/\hbar,
\label{condition2}
\end{equation}
and (\ref{condition1})  are satisfied,  the excited  state of  the
atom  can  be  eliminated  adiabatically  in  analogy  with   Eq.\
(\ref{3.4}).   Making these  and the  rotating wave approximations
one can reduce the Shr\"{o}dinger equation so that it will involve
only   the    wave   functions    $\tilde\psi_{1}(\vc{p},t)$   and
$\tilde\psi_{2}(\vc{p},t)$  of   atomic  motion   in  the   states
$|1\rangle$ and  $|2\rangle$ correspondingly.   When  rewritten in
terms  of  closed  family  wave  functions \cite{Korsunsky,Aspect}
($m=1,2$)
\begin{equation}
\psi_{m}(\vc{p},t)=
\exp[(i/\hbar) E_{|3\rangle}(0)t
-i\omega_{m}t]
\tilde \psi_{m}(\vc{p} - \hbar\vc{k}_{m},t),
\label{4.2}
\end{equation}
this equation takes the form
\begin{mathletters}
\begin{equation}
i\frac{\partial}{\partial t} \psi_{1}(\vc{p},t) =
t_{1}(\vc{p})\psi_{1}(\vc{p},t)
+g_{2}\psi_{2}(\vc{p},t),
\label{4.3a}
\end{equation}
\begin{equation}
i\frac{\partial}{\partial t} \psi_{2}(\vc{p},t) =
[t_{2}(\vc{p})-i\vc{f}_2\cdot\nabla]\psi_{2}(\vc{p},t)
+g_{1}\psi_{1}(\vc{p},t),
\label{4.3b}
\end{equation}
\end{mathletters}
where                 $g_{1}=\Omega^{*}_{2}\Omega_{1}/\Delta_{1}$,
$g_{2}=\Omega^{*}_{1}\Omega_{2}/\Delta_{2}$, and
\begin{equation}
t_{m}(\vc{p})  =
t(\vc{p} - \hbar\vc{k}_{m} ) +\Delta_{m}
+\frac{|\Omega_{m}|^2}{\Delta_{m}}.
\label{4.4}
\end{equation}

\subsection{Quadrature solution}
\label{sec4.1}

Below the exact solution of  the set of equations (\ref{4.3a})  --
(\ref{4.3b}) is presented.   First, the Laplace transformation  is
taken ($m=1,2$)
\begin{equation}
\psi_{m}(\vc{p},\lambda) =
\int_{0}^{\infty} dt e^{-\lambda t}
\psi_{m}(\vc{p},t)
\label{4.5}
\end{equation}
with     the     initial     conditions     $\psi_{m}(\vc{p},t=0)=
\psi_{m}(\vc{p})$.    Second,  a  Cartesian  coordinate  system in
momentum  space  is  introduced  $\vc{p}=(p_x,p_y,p_z)$  with  the
direction  of  z-axis  chosen  opposite  to the vector $\vc{f}_2$.
Then  the  equations  for  the  Laplace transforms read as follows
($f_2=|\vc{f}_2|$):
\begin{mathletters}
\begin{equation}
[t_{1}(\vc{p})- i \lambda]\psi_{1}(\vc{p},\lambda)
+g_2\psi_{2}(\vc{p},\lambda) = -i\psi_{1}(\vc{p}),
\label{4.6a}
\end{equation}
\begin{equation}
\left[t_{2}(\vc{p}) + i f_2 \frac{\partial}{\partial p_z}
- i \lambda \right]\psi_{2}(\vc{p},\lambda)
+ g_{1}\psi_{1}(\vc{p},\lambda) = -i\psi_{2}(\vc{p}).
\label{4.6b}
\end{equation}
\end{mathletters}

Expressing      $\psi_{1}(\vc{p},\lambda)$      via      $\psi_{2}
(\vc{p},\lambda)$ from the Eq.\ (\ref{4.6a})
\begin{equation}
\psi_{1}(\vc{p},\lambda) =
\frac{-g_2\psi_{2}(\vc{p},\lambda)-i\psi_{1}(\vc{p})}
{t_{1}(\vc{p})- i \lambda}
\label{4.7}
\end{equation}
one can  simplify Eq.\  (\ref{4.6b}) so  that it  will contain the
only unknown function $\psi_{2}(\vc{p},\lambda)$ and will allow an
easy solution  after imposing  an appropriate  boundary condition.
This condition may be found if  we take into account that an  atom
initially having a finite z-component  of momentum is not able  to
reach infinitely large positive $p_z$  at any time because of  the
force $f_2$ acting in opposite direction, i.e., it is necessary to
put
\begin{equation}
\psi_{2}(\vc{p},\lambda)|_{p_z\to \infty}=0.
\label{4.8}
\end{equation}
In such a way one gets
\begin{eqnarray}
\psi_{2}(\vc{p},\lambda) = &&
\frac{1}{f_2} \int^{\infty}_{p_z} d p'_z
\left[\psi_{2}(\vc{p}') -
\frac{g_1 \psi_{1}(\vc{p}')}
{t_{1}(\vc{p}')- i \lambda} \right]
\nonumber\\
&&\times
\exp\left\{ \frac{i}{f_2}
[h(\vc{p},\vc{p}',\lambda)
-i\lambda (p_z - p'_z)]
\right\},
\label{4.9}
\end{eqnarray}
where $p_x$ and  $p_y$ are fixed,  i.e., $\vc{p}'=(p_x,p_y,p'_z)$,
and
\begin{equation}
h(\vc{p},\vc{p}',\lambda) =
\int^{p_z}_{p'_z} d p''_z
\left[t_{2}(\vc{p}'')
-\frac{g_1 g_2}
{t_{1}(\vc{p}'')- i \lambda} \right].
\label{4.10}
\end{equation}

The  inverse  Laplace  transformation  of  expressions (\ref{4.7})
(\ref{4.9})  with   the  Mellin   formula  produces   the  desired
quadrature solution ($m=1,2$)
\begin{equation}
\psi_{m}(\vc{p},t) = 2\pi i
\int_{\epsilon-i\infty}^{\epsilon+i\infty}
d\lambda  e^{\lambda t}
\psi_{m}(\vc{p},\lambda),
\quad \epsilon > 0.
\label{4.11}
\end{equation}

\subsection{Approximate formulas}
\label{sec4.2}

Although the  integral in  Eq.\ (\ref{4.11})  cannot be calculated
explicitly, it becomes possible to tabulate it after employing the
following approximation.  Note that by the retardation theorem the
value of $\psi_{\{1,2\}}(\vc{p},t)$ is nonzero only if $t+ (p_z  -
p'_z)/f_2 >0$, which has a clear physical interpretation:  due  to
the action  of force  $f_2$ a  state with  the momentum  $p_z$ may
arise only from  the states which  span the interval  $p_z< p'_z <
p_z + f_2 t$.  At times considered here ($\sim$ time of the  $\pi$
pulse) this interval appears to  be narrow in comparison with  the
typical momentum in  the system $f_2  t \ll \Delta_p$.   Therefore
one can expand the function $h(\vc{p},\vc{p}',\lambda)$ as a power
series in $p_z - p'_z$ and retain only the linear term
\begin{equation}
h(\vc{p},\vc{p}',\lambda) \simeq
(p_z- p'_z)
\left[t_{2}(\vc{p})
-\frac{g_1 g_2}
{t_{1}(\vc{p})- i \lambda} \right].
\label{4.12}
\end{equation}
The  validity  of  such  an  approximation  is  restricted  by the
condition
\begin{equation}
t \ll \sqrt{\pi\hbar M/(f_2 \Delta_p)}
\label{condition3}
\end{equation}
preserving  the  phase  of  integrand  in Eq.\ (\ref{4.9}) against
considerable    variation    due     to    omitted    terms     in
$h(\vc{p},\vc{p}',\lambda)$.  In  the case of  Na it requires  the
interaction time to be much less than $3.8\times 10^{-3}$ c.

After  changing  the  order  of  integration  over $d p'_z$ and $d
\lambda$,  subsequent  calculations  are  obvious  and lead to the
following  expressions  for  the  Green function components ($m,n=
1,2$):
\begin{eqnarray}
G_{mn}(\vc{p},\vc{p}',t) =&&
D_{mn}(\vc{p}+\hbar\vc{k}_{m},\vc{p}'+\hbar\vc{k}_{n},t)
\nonumber\\
&&\times
\delta_{\perp}[\vc{p} +
\hbar(\vc{k}_{m} - \vc{k}_{n})- \vc{p}']
\nonumber\\
&&\times
\exp[i\omega_{m}t
-(i/\hbar)  E_{|3\rangle}(0)t],
\label{4.13}
\end{eqnarray}
which   determine   the   evolution   of  wave  functions  $\tilde
\psi_{\{1,2\}}(\vc{p},t)$.    Here  $\delta_{\perp}(\vc{p}) \equiv
\delta(p_x) \delta(p_z)$ and
\begin{mathletters}
\begin{eqnarray}
D_{11}(\vc{p},\vc{p}',t) = &&
\exp[-i t_{1}(\vc{p}) t]
\delta(p_z-p'_z)
\nonumber\\
&&- \sqrt{g_1 g_2} \eta(\vc{p},\vc{p}',t)
\chi_{(+)}(p_z-p'_z,t),
\label{4.14a}
\end{eqnarray}
\begin{eqnarray}
D_{22}(\vc{p},\vc{p}',t) = &&
\exp[-i t_{2}(\vc{p}) t]
\delta(p_z+ f_2 t - p'_z)
\nonumber\\
&&- \sqrt{g_1 g_2}
\eta(\vc{p},\vc{p}',t)
\chi_{(-)}(p_z-p'_z,t),
\label{4.14b}
\end{eqnarray}
\begin{equation}
D_{12}(\vc{p},\vc{p}',t) =
- i g_2 \eta(\vc{p},\vc{p}',t)
\chi_{(0)}(p_z-p'_z,t),
\label{4.14c}
\end{equation}
\begin{equation}
D_{21}(\vc{p},\vc{p}',t) =
- i g_1 \eta(\vc{p},\vc{p}',t)
\chi_{(0)}(p_z-p'_z,t).
\label{4.14d}
\end{equation}
\end{mathletters}
The   values   $\eta(\vc{p},\vc{p}',t)$   and  $\chi_{(k)}(p_z,t)$
($k=0,\pm$)  used  in  these  equations contain the theta-function
$\theta()$  and  the  Bessel  functions  of the first kind $J_n()$
($n=0,1$)
\begin{eqnarray*}
\eta(\vc{p},\vc{p}',t) = &&
\exp\left\{ \frac{i}{f_2}(p_z - p'_z)
[t_{2}(\vc{p}) - t_{1}(\vc{p})]
-i t_{1}(\vc{p}) t \right\}
\nonumber \\
&& \times
\frac{1}{f_2}
\theta(p_z+ f_2 t - p'_z)
\theta(p'_z - p_z),
\end{eqnarray*}
\[
\chi_{(0)}(p_z,t) =
J_0\left\{
t \sqrt{g_1 g_2 [1-f^{2}(p_z,t)]}
\right\},
\]
\begin{eqnarray*}
\chi_{(\pm)}(p_z,t) =&&
\frac{1 \pm f(p_z,t)}
{\sqrt{1-f^{2}(p_z,t)}}
\nonumber \\
&&\times J_1\left\{
t \sqrt{g_1 g_2 [1-f^{2}(p_z,t)]}
\right\},
\end{eqnarray*}
where $f(p_z,t) = 1+ 2p_z/(f_2 t)$.

\subsection{Velocity  selectivity}
\label{sec4.3}

To involve all  atoms into the  DS filling one  must use a  set of
Raman pulses during every cooling cycle adjusting their number and
the difference of  beam frequencies for  each pulse repetition  in
accordance with the  velocity selectivity of  discussed transition
$|1\rangle \to |2\rangle$, which appears to depend, besides  other
factors, on  the size  ($\sim 2L$)  of the  atomic sample or, more
precisely, on  the width  of the  initial coordinate distribution.
Indeed, any  finite coordinate  distribution, if  fitted with  the
Gaussian profile $\propto \exp(-\vc{r}^2/ L^{2})$, gives rise to a
factor $\exp[-L^{2} (\vc{p}'_{1} - \vc{p}'_{2})^{2}/ (4 \hbar^2)]$
in the density matrix $\rho_{11} (\vc{p}'_1, \vc{p}'_2,0)$.   When
$(L f_2 t/\hbar)^2 \ll 1$ (a thin atomic sample), this exponential
factor   does    not   influence    the   integration    over   $d
\vc{p}'_{\{1,2\}}$ in  Eq.\ (\ref{3.2}),  and for  a flat  initial
atomic momentum distribution one gets standard formulas describing
velocity-selective transitions in a three-level system without  an
external potential (see, e.g.,  \cite{Korsunsky}).  In this  case,
after $\pi$-pulse time
\begin{equation}
\tau_0 = \pi/\left( 2\sqrt{g_1 g_2} \right)
\label{4.19}
\end{equation}
the width of selection (half-width at $1/e$ of maximum) becomes
\begin{equation}
\delta^{(1)} p \simeq \hbar \sqrt{g_1 g_2} M/\Delta_p.
\label{4.20}
\end{equation}
Conversely,  at  large  $L$  (when  one  can  put $\vc{p}_1 \simeq
\vc{p}_2  \simeq  \vc{p}$)  the  damping  of the integrand in Eq.\
(\ref{3.2})  is  more  rapid  than  phase  variation  of the Green
functions   product   $\propto    \exp   \{i   [t_{2}(\vc{p})    -
t_{1}(\vc{p})]  (p'_{1z}  -  p'_{2z})/f_2\}$,  and  the   velocity
selectivity, which at small $L$ has been provided by the resonance
denominator, disappears.  Instead, if $\hbar/L \ll f_2 t$, in  the
closed family  basis it  is determined  by the  exponential factor
$\sim  \exp  \{-\hbar^2  [t_{1}(\vc{p})  -  t_{2}(\vc{p})]^2/ (L^2
f_2^2) \}$ because  of negligible theta-function  contributions to
the integral.  Such picture is in agreement with Eq.\ (\ref{eq5}).
Thus,  the  width  of  selection  may  be  evaluated as [cf.  Eq.\
(\ref{tzwidth})]
\begin{equation}
\delta^{(2)} p \simeq M L f_2/(2\Delta_p).
\label{4.21}
\end{equation}

To illustrate this point  in a 1D configuration  (when $\vc{k}_{1}
\uparrow  \downarrow  \vc{k}_{2}$  and  $\vc{\Delta}_{p}  \uparrow
\downarrow  \vc{f}_2$)  let  us  regard  a population $\rho_{22} =
\rho_{22}(p,p,\tau_0)$ of the state $|2 \rangle$ as a function  of
the  momentum  component  $p=p_z$  and the dimensionless parameter
$\vartheta =  L f_2  \tau_0/\hbar$.   Figure \ref{fig2}  shows the
corresponding  dependence  calculated  for  sodium  atoms with the
initial density matrix  $\rho_{11}(p_1,p_2,0)= \exp[-L^{2} (p_1  -
p_2)^{2}/(4  \hbar^2)]$   after  $\pi$-pulse   time.     The  peak
characterizing  velocity  selectivity  of  the  Raman   excitation
decreases with  increasing $\vartheta$,  and begins  to widen from
the point  $\vartheta =  \pi$ where  $\delta^{(1)} p =\delta^{(2)}
p$.  So  the parameter $\vartheta$  discriminates between the  two
domains with  different behaviors  of the  selection width:   when
$\vartheta \ll \pi$ it is given by Eq.\ (\ref{4.20}) and does  not
depend on the size of the atomic sample, whereas at $\vartheta \gg
\pi$ it broadens  with $L$ in  accordance with Eq.\  (\ref{4.21}).
In any case  however, the width  of selection cannot  be less than
$\delta^{(2)} p$.

\section{OPTICAL PUMPING}
\label{sec5}

In this section,  the optical-pumping beam  is considered to  be a
plane electromagnetic wave (\ref{3.1}) at $m=2$, whose interaction
with atomic sample is described by means of a master equation  for
the   density   matrix   in   external   potential   (see,   e.g.,
\cite{Soroko}).  In the closed family basis \cite{Aspect,Matisov},
where  the  density  matrix  is  denoted  $\sigma_{ab}(\vc{p}_{1},
\vc{p}_{2},t)$, $a,b  = 1,2,3$,  and employing  the rotating  wave
approximation, the master equation may be rewritten as
\begin{eqnarray}
i\frac{\partial}{\partial t}
\sigma_{ab}(\vc{p}_{1},\vc{p}_{2},t) =&&
\sum_{cd} \left[
H_{ac}(\vc{p}_{1})
\delta_{bd} -
H^{*}_{db}(\vc{p}_{2})
\delta_{ac}
\right]
\nonumber\\
&&\times
\sigma_{cd}(\vc{p}_{1},\vc{p}_{2},t)
+ {\cal R}_{ab}[\sigma].
\label{5.1}
\end{eqnarray}
The Hamiltonian $H_{ac}(\vc{p})$ is non-Hermitian because it  governs
the damping due to spontaneous decay of the excited state
\begin{eqnarray}
H_{ab}(\vc{p}) =&&
\left[ t(\vc{p}-\delta_{a 2}\hbar\vc{k}_2)
+ \Delta_{a}
-i(\delta_{a 2}\vc{f}_2 + \delta_{a 3}\vc{f}_3)
\cdot\nabla \right] \delta_{ab}
\nonumber\\
&&- i(\gamma/2) \delta_{a 3}\delta_{b 3}
-\Omega_2  \delta_{a 3}\delta_{b 2}.
\label{5.2}
\end{eqnarray}
The last term in Eq.\ (\ref{5.1}) is responsible for the return of
the atom to the ground state via spontaneous emission
\begin{eqnarray}
{\cal R}_{ab}[\sigma] = &&
\delta_{a 1} \delta_{b 1} \gamma
\int d \hat\vc{n} \Phi(\hat\vc{n})
\nonumber\\
&&\times
\sigma_{33}(\vc{p}_{1} -\hbar k_{31}\hat\vc{n},
\vc{p}_{2} - \hbar k_{31}\hat\vc{n}, t),
\label{5.3}
\end{eqnarray}
where  the  function  $\Phi(\hat\vc{n})$  determines  the relative
probability   of   emitting   a   photon   with  the  wave  number
$k_{31}=[E_{|3\rangle}(0)  -  E_{|1\rangle}(0)]/(\hbar  c)$ in the
$\hat\vc{n}$   direction.      Below,   the   spherical   symmetry
approximation     is      adopted     for      $\Phi(\hat\vc{n})$:
$\Phi(\hat\vc{n}) \simeq 1/(4\pi)$.

Note that independent of  the shape of excited-state  distribution
$\sigma_{33}$, the profile  of atoms repumped  into the state  $|1
\rangle$   is   represented   by   a   smooth   functional  ${\cal
R}_{ab}[\sigma]$ which  undergoes sufficient  variation only  when
its momentum  argument changes  by $\sim  \hbar k_{31}$  as it  is
evident from Eq.\ (\ref{5.3}).   A powerful simplification  of the
master  equation  may  be  achieved  if  we take into account that
neither  of  the  forces  $\vc{f}_2$  and  $\vc{f}_3$  affect this
functional  significantly  when  the  spatial  shift of transition
frequency is small enough
\begin{equation}
\max(|\Delta_{2}|, \gamma/2) \gg
|\vc{f}_3-\vc{f}_2|L/\hbar.
\label{condition4}
\end{equation}
Indeed, the last  condition protects the  excited-level population
from the influence  of the external  potential, and the  effect of
forces reduces to a momentum kick $\delta \vc{p}_{k} \sim \vc{f}_2
\tau_p + \vc{f}_3 \tau_d$ received by an atom, where $\tau_p$  and
$\tau_d  \simeq  1/\gamma$  are  the  lifetimes  in the states $|2
\rangle$ and $|3 \rangle$ respectively.  Evaluating $\tau_p$  from
the optical pump rate (see, e.g., \cite{Pellizzari,Letokhov}) as
\begin{equation}
\tau_p \simeq
\frac{\gamma^2 + 4\Delta_{2}^2}
{4|\Omega_2|^{2}\gamma},
\end{equation}
one can find that for a  realistic set of parameters the value  of
the kick appears to be too small to induce a noticeable  variation
in the distribution of  repumped atoms:  $|\delta  \vc{p}_{k}| \ll
\hbar  k_{31}$.    For  example,  in  the  case of sodium $|\delta
\vc{p}_{k}|   \simeq    5\times   10^{-6}    \hbar   k_{31}$    if
$|\Omega_2|=0.1\gamma$,   and   $\Delta_{2}=0$.     Therefore  the
potential energy term containing forces $\vc{f}_2$ and  $\vc{f}_3$
can be omitted in the Hamiltonian (\ref{5.2}).

After  such  a  simplification  the  master  equation  admits   an
analytical solution, which one can obtain, e.g., with the help  of
Laplace  transform.    As  a  result,  the  only  component of the
original-basis  density  matrix,  nonvanishing  at $t \gg \tau_p$,
takes the form
\begin{eqnarray}
\rho_{11}(\vc{p}_{1},\vc{p}_{2},t) = &&
\rho_{11}(\vc{p}_{1},\vc{p}_{2},0)
\exp(-i\lambda_0 t)
-\frac{i\gamma |\Omega_2|^2}{4\pi}
\int \frac{d \hat\vc{n}}
{\sqrt{D(\vc{p}'_1)}
\left[\sqrt{D(\vc{p}'_2)}\right]^*}
\sum_{\alpha = \pm 1}
\sum_{\beta = \pm 1}
(-1)^{\alpha + \beta}
\nonumber\\
&&\times
\frac{\exp(-i\lambda_0 t) -
\exp\{-i t[\lambda_{(\alpha)}(\vc{p}'_1)
- \lambda^{*}_{(\beta)}(\vc{p}'_2)]\}}
{\lambda_{(\alpha)}(\vc{p}'_1)
- \lambda^{*}_{(\beta)}(\vc{p}'_2)
-\lambda_0}
\rho_{22}(\vc{p}'_{1} + \hbar\vc{k}_2,
\vc{p}'_{2}+ \hbar\vc{k}_2,0).
\label{5.6}
\end{eqnarray}
In  this  equation  $\vc{p}'_{\{1,2\}}  =  \vc{p}_{\{1,2\}}- \hbar
k_{31}\hat\vc{n}$, and
\[
\lambda_0 = t(\vc{p}_1) - t(\vc{p}_2),
\]
\[
D(\vc{p}) = [\delta^{(-)}(\vc{p})]^2 + 4|\Omega_2|^2,
\]
\[
\lambda_{(\pm 1)} = \frac{1}{2}
\left[\delta^{(+)}(\vc{p}) \pm \sqrt{D(\vc{p})}\right],
\]
where
\[
\delta^{(\pm)}(\vc{p})=
t(\vc{p} - \hbar\vc{k}_2) +\Delta_2 \pm
[t(\vc{p}) -i\gamma/2].
\]

The above solution is valid when the spontaneous decay rates  into
states  other  than  $|1  \rangle$  are  negligible.    Due to the
specific choice of $|3 \rangle$  as the working excited state  all
such decay rates turn out to be $\eta^2_{\rm HF} \ll 1$ times less
than $\gamma$.  Nevertheless, in real atomic system some  fraction
of atoms will accumulate in unwanted states.  To return them  back
into the ground level one should include additional laser beams in
the  considered  scheme,  as  it  is  done,  e.g., in the coherent
optical pumping \cite{Korsunsky2}.

Another complication may arise from non-resonance excitations  out
of the ground level which we  do not include in the treatment  for
both  the  {\it  second}  and  {\it  third  steps} of cooling.  In
effect, the laser  light driving the  $|2 \rangle \to  |3 \rangle$
transition introduces  a detuning  $\sim \omega_{\rm  HFS}$ and  a
Rabi frequency $\sim \Omega_2/ \eta_{\rm HF}$ with respect to  the
transition $|1 \rangle \to |3  \rangle$.  However, if we  restrict
$\Omega_2$ by the condition
\begin{equation}
\Omega_2 \ll \omega_{\rm  HFS} \eta_{\rm  HF},
\label{5.7}
\end{equation}
the excited-state population  [$\propto \Omega^2_2 /  (\omega_{\rm
HFS} \eta_{\rm HF})^2$] appears to be small.

\section{NUMERICAL SIMULATION}
\label{sec6}

In the following we  present one-dimensional results obtained  for
Na assuming that all vectors have only z-components, i.e., lie  on
the same axis  with the gravitational  force, and the  laser beams
with $\vc{k}_1$ and $\vc{k}_2$ are counterpropagating.  An initial
distribution of ground-state atoms is considered to be Gaussian
\begin{equation}
\rho_{11}(p_1,p_2,0) =
\frac{L}
{\hbar\pi^{1/2}\sqrt{1+2L^2\sigma_p^2}}
\exp\left\{
\frac{-L^2}{4 \hbar^2}
\left[\frac{(p_1 + p_2)^{2}}
{1+2L^2\sigma_p^2}
+(p_1 - p_2)^{2}\right]
\right\},
\label{6.1}
\end{equation}
where $p=p_z$, and $\sigma_p$  stands for momentum dispersion  (in
units of $\hbar$).   Since in our scheme  we imply that an  atomic
sample precooled to the recoil limit is used, it is reasonable  to
take  the  wave  number  of  laser  light  $k  = 1.07 \times 10^5$
cm$^{-1}$  as  an  input  for  $\sigma_p$.    Note  that  for  the
considered laser-beams geometry $\Delta_p/\hbar = k$.  In order to
satisfy  the  condition  (\ref{5.7})  we  also  take the parameter
$\eta_{\rm HF} = 2.5 \times 10^{-2}$, which corresponds to $B_0  =
10^3$ G.

\begin{figure}
\epsfbox{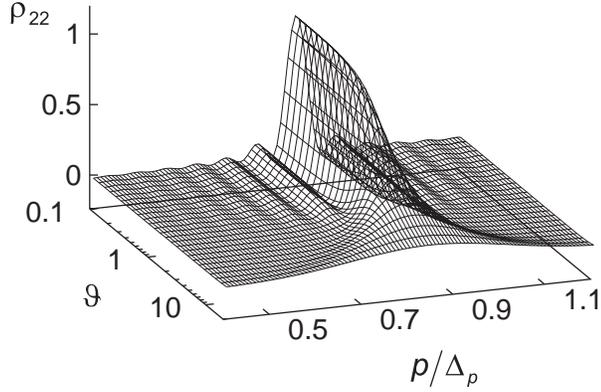}
\caption{
Velocity  selectivity  of  Raman  excitation  in  1D  case  for Na
represented  as  the  dependence  of  $\rho_{22}$,  the  state $|2
\rangle$  population,  on  the  momentum  component  $p=p_z$   and
parameter $\vartheta = L f_2 \tau_0/\hbar$.  Detunings $\Delta_1 =
\Delta_2  =   -  16\gamma$   and  Rabi   frequencies  $\Omega_1  =
0.16\gamma$, $\Omega_2 = 0.016\gamma$.
}
\label{fig2}
\end{figure}

In  the  numerical  simulation  of  a  cooling  cycle  each   {\it
first-step}  pulse  was  followed  by  five  repetitions  of a set
involving seven Raman and one  optical pumping pulses.  A  mapping
between the density matrices at the beginning and the end of  this
sequence  of  pulses  is   given  by  the  formulas   (\ref{3.2}),
(\ref{3.10}), (\ref{4.13}), and (\ref{5.6}).  Both {\it first} and
{\it  second  steps}  of  cooling  continued during $\tau_0$-times
defined  according  to  the  Eqs.\  (\ref{3.13})  and (\ref{4.19})
correspondingly.  The  duration of the  optical pumping pulse  was
taken to be $10 \tau_p$ to provide a complete depopulation of  the
$|2  \rangle$  state.    The  remaining  parameters were chosen as
follows.  For the {\it first  step}:  $\Omega_1 = \Omega_2 =  0.08
\gamma$, and $\Delta_1 = - 16\gamma$.  For the {\it second  step}:
$\Omega_1 = 0.4 \gamma$, $\Omega_2  = 0.04 \gamma$, and all  seven
Raman pulses were detuned to the  red so that the sum $\Delta_1  +
\Delta_2  =  -  32\gamma$  remained  constant while the difference
$\Delta_1 - \Delta_2$ was increased  by -135, 118, 372, 625,  880,
1135, and 1393 kHz.  Such a choice of detunings was tailored  both
to  span  the  momentum  interval  $0\le  p  \le 3\Delta_p$ and to
minimize the losses of  DS population due to  parasitic excitation
by sidelobes  in the  frequency spectrum  of Raman  transitions at
$\vartheta <  \pi$ (see  Fig.\ \ref{fig2}).   For  the {\it  third
step}  we  put  $\Omega_2  =  0.1\gamma$,  and  $\Delta_2=0$.  The
initial size of atomic sample was taken $L = 1$ cm.  However,  for
the  given  set  of  Raman  light  parameters  this, or indeed any
smaller, value of $L$ leads  to the inequality $\vartheta <  \pi$.
It means  that the  width of  velocity selection  is determined by
Eq.\  (\ref{4.20})  and  does  not  depend  on $L$.  Therefore our
results remain correct for all $L\le 1$ cm.

\begin{figure}
\epsfbox{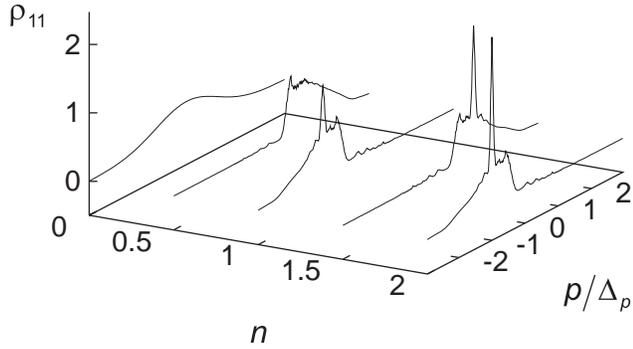}
\caption{
One-dimensional ground-state distribution $\rho_{11}$ for Na as  a
function of  atomic momentum  $p$ and  the number  $n$ of  cooling
cycles   normalized   to   $1$   on   the   scale  $p/\Delta_{p}$.
Half-integer  values  of  $n$  correspond  to the beginning of the
second stage of each cooling  cycle which starts when laser  beams
reverse.   The curve  with $n=0$  gives the  initial distribution.
The highest peaks of the function represent the DS.
}
\label{fig3}
\end{figure}

\begin{figure}
\epsfbox{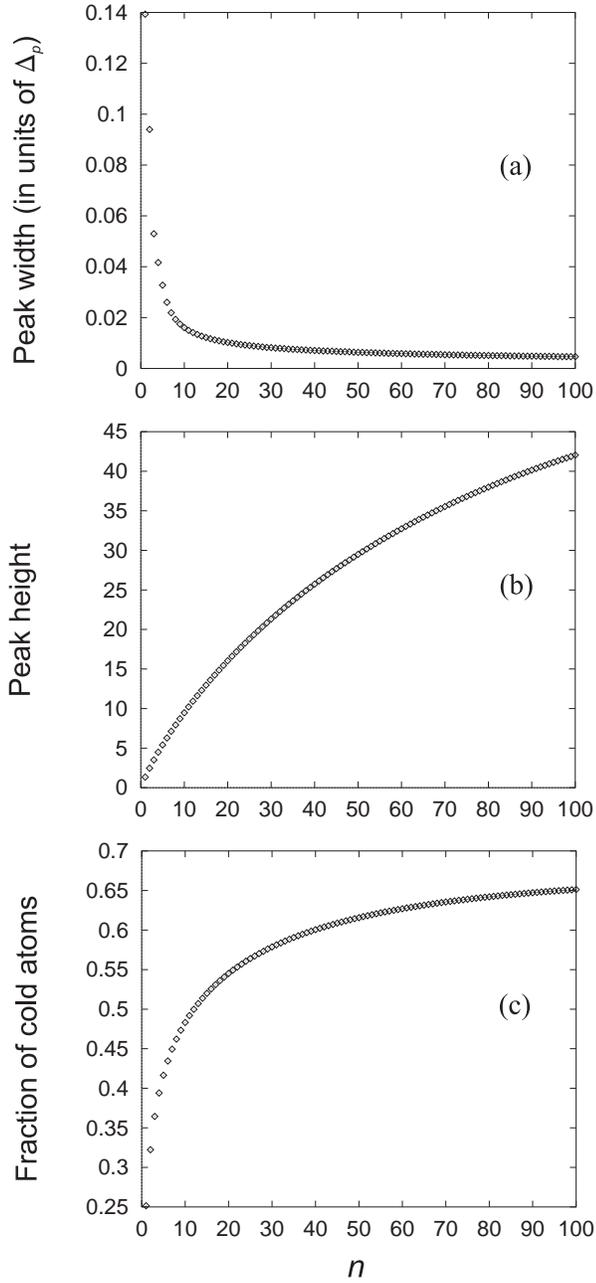}
\caption{
Parameters of the DS  peak vs the number  of cooling cycles.   (a)
Peak width at the half-maximum, (b) peak height, and (c)  fraction
of cold atoms around DS in the interval $\pm \delta p \approx  \pm
0.14  \Delta_{p}$,  which  spreads  between  the  first  minima in
momentum dependence of the $|1 \rangle \to |1 \rangle$  transition
rate.
}
\label{fig4}
\end{figure}

Figure  \ref{fig3}   shows  the   initial  momentum   distribution
$\rho_{11}(p,p,0)$ and the formation of a DS peak during two first
cooling cycles including intermediate stages when the position  of
this  peak  is  alternated.    Although  each  $|1  \rangle \to |1
\rangle$  transition  captures  atoms  in  a  rather wide momentum
interval $\sim 2\delta p \approx 0.28 \Delta_p$, the width of  the
DS peak  (at half-maximum)  decreases rapidly  with the  number of
applied cycles because of  a pronounced maximum in  the transition
rate  profile.    After  10  cycles  the  decrease  slows down and
approaches at  $0.005\Delta_p$ by  the end  of cooling,  as may be
seen from Fig.\ \ref{fig4}(a).  At the same time, the peak  height
growth is far from saturation, as depicted in Fig.\ \ref{fig4}(b).
After 100  cooling cycles  the height  is more  than 98  times the
initial distribution maximum.

The fraction of cold atoms in the interval $-\Delta_p-\delta p \le
p \le -\Delta_p+\delta  p$ depends on  the difference between  the
feeding rate due to optical pumping and losses during DS transfer,
which arise along  with the reduction  of the peak  width.  Figure
\ref{fig4}(c) shows that about 65 \% of all atoms collect there by
the end of cooling.

\begin{figure}
\epsfbox{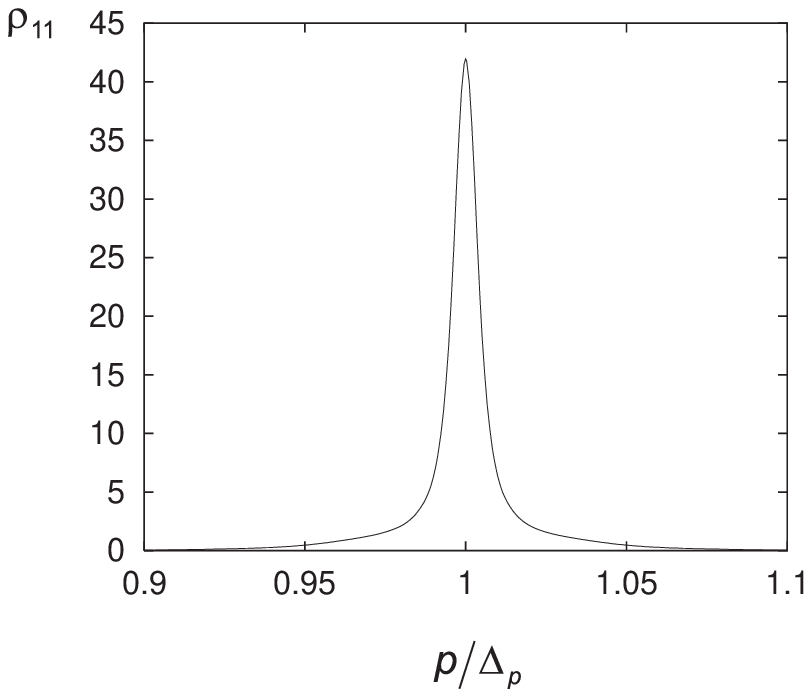}
\caption[Figure 5]{
Final ground-state distribution $\rho_{11}$ of atoms with positive
momenta as  a function  of $p$.   The  full-scale distribution  is
normalized as  in the  Fig.\ \ref{fig3}  whereas the  plotted part
contains $\sim 59\%$ of all particles.
}
\label{fig5}
\end{figure}

When separated from the  background by the final  {\it first-step}
transition which transfers the aforementioned interval to positive
momentum half-axis, the  DS peak acquires  a shape represented  in
the  Fig.\  \ref{fig5}.    As  a result, the effective temperature
calculated  as  a  mean  kinetic  energy  of the atoms distributed
within the domain $\Delta_p- \delta  p \le p \le \Delta_p+  \delta
p$ reaches $0.4$ nK or $0.015 T_G$.

A  very  special  sequence  of  laser  pulses  considered  in  our
numerical simulation was designed  both to minimize the  volume of
computations and to demonstrate  the possibility of cooling  below
the gravitational  limit with  a noticeable  efficiency.  However,
this sequence is not the best from the point of view of  practical
application  because  it  includes  Raman  pulses  in  the  regime
$\vartheta  <  \pi$,  in  which  excitation-spectrum sidelobes can
destroy  DS   unless  one   correctly  adjusts   detunings.    For
experimental purposes the alternative choice $\vartheta > \pi$ may
be much more attractive insofar  as it does not lead  to parasitic
excitations, as illustrated  in Fig.\ \ref{fig2}.   Unfortunately,
in this case each cooling cycle must contain a considerable number
of  steps  because  of  the  moderate  excitation  rate  of  Raman
transition, and a numerical simulation does not seem feasible.

\section{CONCLUSIONS}
\label{sec7}

In this  paper we  have studied  a modification  of Raman  cooling
method,  in  which  the  atomic  internal ground state possesses a
translational invariance  due to  compensation of  gravity by  the
Stern-Gerlach effect.   Our  scheme is  based on  creating a  dark
state which is cyclically moved in momentum space with  additional
velocity-selective  two-photon  transitions  so  that  it  is kept
unreachable for Raman excitation-repumping pulse sequences at  all
times.   The consideration  has been  restricted to  dilute atomic
samples,  i.e.  we  have  not  included any many-atom interactions
\cite{Lewenstein}.  In  the approximation in  which the DS  losses
induced  by  atomic  collisions,  non-resonance  excitations,  and
photon  scattering  are  neglected,  our one-dimensional numerical
computations  have  shown  that  a  considerable  fraction  of all
particles can be cooled to the temperature below the gravitational
limit.      Furthermore,   this   temperature   can  be  decreased
significantly for laser  pulses which provide  a smaller width  of
the velocity selection during DS transfer.  Though we have given a
numerical   example   only   for   a  one-dimensional  model,  our
theoretical investigation  of the  suggested scheme  is also valid
for the two- and three-dimensional cases.

In a  bosonic system,  where the  losses in  DS population  can be
compensated  by  the  quantum-statistical  enhancement  of feeding
rate, our scheme  will be appropriate  for dense samples  as well.
Thus,  it  may  be  used  in  creation  of  a coherent atomic-beam
generator \cite{Guzman,CAB}.  An  easy tunable wavelength will  be
one of the advantages of such a device, because the momentum of  a
cooled atom is readily defined by the geometry of laser beams.

% now the references
%

\input{epsf.tex}


\begin{references}
%
\bibitem{Phillips}P. Lett, R. Watts, C. Westbrook, W. Phillips,
P. Gould, and H. Metcalf, Phys. Rev. Lett. {\bf 61}, 169 (1988).
%
\bibitem{VSCPT}A. Aspect, E. Arimondo, R. Kaiser, N.
Vansteenkiste, and C. Cohen-Tannoudji,
Phys. Rev. Lett. {\bf 61}, 826 (1988); J. Lawall, S. Kulin, B.
Saubamea, N. Bigelow, M. Leduc, and C. Cohen-Tannoudji, {\it
ibid}. {\bf 75}, 4194 (1995).
%
\bibitem{Raman}M. Kasevich and S. Chu, Phys. Rev. Lett. {\bf
69}, 1741 (1992); N. Davidson, H. J. Lee, M. Kasevich, and S. Chu,
{\it ibid}. {\bf 72}, 3158 (1994); H. J. Lee, C. S. Adams, M.
Kasevich, and S. Chu, {\it ibid}. {\bf 76}, 2658 (1996).
%
\bibitem{Dum}R. Dum, Phys. Rev. A {\bf 54}, 3299 (1996).
%
\bibitem{Dum2}R. Dum and M. Ol'shanii, Phys. Rev. A {\bf 55}, 1217
(1997).
%
\bibitem{Pellizzari}T. Pellizzari, P. Marte, and P. Zoller,
Phys. Rev. A {\bf 52}, 4709 (1995).
%
\bibitem{Morigi}G. Morigi, J. I. Cirac, K. Ellinger, and P.
Zoller, Phys. Rev. A {\bf 57}, 2209 (1998).
%
\bibitem{Reichel}J. Reichel, F. Bardou, M. Ben Dahan, E. Peik, S.
Rand, C. Salomon, and C. Cohen-Tannoudji,
Phys. Rev. Lett. {\bf 75}, 4575 (1995).
%
\bibitem{Moskowitz}P. E. Moskowitz, P. L. Gould, S. R. Atlas, and
D. E. Pritchard, Phys. Rev. Lett. {\bf 51}, 370 (1983).
%
\bibitem{Martin}P. J. Martin, B. C. Oldaker, A. N. Miklich, and
D. E. Pritchard, Phys. Rev. Lett. {\bf 60}, 515 (1988).
%
\bibitem{Soroko}A. V. Soroko, J. Phys. B. {\bf 30}, 5621 (1997).
%
\bibitem{Kazantsev}A. P. Kazantsev, G. A. Ryabenko, G. I.
Surdutovich, and V. P. Yakovlev, Phys. Rep. {\bf129}, 75 (1985).
%
\bibitem{Zhang2}W. Zhang and D. F. Walls,  Phys. Rev. A {\bf 52},
4696 (1995).
%
\bibitem{Korsunsky}E. A. Korsunsky, D. V. Kosachiov, B. G.
Matisov, and Yu. V. Rozhdestvensky, JETP {\bf 76}, 210 (1993).
%
\bibitem{Zhang}W. Zhang and D. F. Walls,  Phys. Rev. A {\bf 49},
3799 (1994).
%
\bibitem{Guzman}A. M. Guzman, M. Moore, and P. Meystre,  Phys.
Rev. A {\bf 53}, 977 (1996).
%
\bibitem{Letokhov}V. Minogin and V. Letokhov, {\it Laser Light
Pressure on Atoms} (Gordon and Breach, New York, 1986).
%
\bibitem{Aspect}A. Aspect, E. Arimondo, R. Kaiser, N.
Vansteenkiste, and C. Cohen-Tannoudji, J. Opt. Soc. Am. B {\bf 6},
2112 (1989).
%
\bibitem{Matisov}B. Matisov, V. Gordienko, E. Korsunsky, and L.
Windholz,  JETP {\bf 80}, 386 (1995).
%
\bibitem{Korsunsky2}E. A. Korsunsky, W. Maichen, and L. Windholz,
Phys. Rev. A {\bf 56}, 3908 (1997).
%
\bibitem{Lewenstein}M. Lewenstein, L. You, J. Cooper, and
K. Burnett, Phys. Rev. A {\bf 50}, 2207 (1994).
%
\bibitem{CAB}R. J. C. Spreeuw, T. Pfau, U. Janicke, and M.
Wilkens, Europhys. Lett. {\bf 32}, 469 (1995); H. M. Wiseman and
M. J. Collett, Phys. Lett. A {\bf 202}, 246 (1995);
M. Holland, K. Burnett, C. Gardiner, J. I. Cirac, and P. Zoller,
Phys. Rev. A {\bf 54}, R1757 (1996); G. M. Moy, J. J. Hope, and C.
M. Savage, {\it ibid}. {\bf 55}, 3631 (1997).
%
\end{references}
\end{document}